\documentclass[preprint]{aastex63}

\usepackage{amsmath,amssymb}
\usepackage{graphicx}
\received{April 11, 2019}
\revised{July 4, 2019}
\accepted{July 18, 2019}
\submitjournal{Advances in Astronomy}

\shorttitle{Time Stamp Recognition in Solar Scanned Images}
\shortauthors{Zhang et al.}

\begin{document}

\title{Intelligent Recognition of Time Stamp Characters in Solar Scanned Images from Film}

\correspondingauthor{Shuguang Zeng}
\email{zengshuguang19@163.com}

\author[0000-0002-3883-548X]{Jiafeng Zhang}
\affiliation{College of Science, China Three Gorges University, Yichang 443002, China}

\author{Guangzhong Lin}
\affiliation{College of Science, China Three Gorges University, Yichang 443002, China}

\author[0000-0002-3216-7476]{Shuguang Zeng}
\affiliation{College of Science, China Three Gorges University, Yichang 443002, China}

\author{Sheng Zheng}
\affiliation{College of Science, China Three Gorges University, Yichang 443002, China}

\author[0000-0003-1675-1995]{Xiao Yang}
\affiliation{Key Laboratory of Solar Activity, National Astronomical Observatories, Chinese Academy of Sciences, Beijing 100101, China}

\author{Ganghua Lin}
\affiliation{Key Laboratory of Solar Activity, National Astronomical Observatories, Chinese Academy of Sciences, Beijing 100101, China}

\author{Xiangyun Zeng}
\affiliation{College of Science, China Three Gorges University, Yichang 443002, China}

\author[0000-0002-5233-565X]{Haimin Wang}
\affiliation{Institute for Space Weather Sciences, New Jersey Institute of Technology, 323 Martin Luther King Boulevard, Newark, NJ 07102-1982, USA}

\begin{abstract}

Prior to the availability of digital cameras, the solar observational images are typically recorded on films, and the information such as date and time were stamped in the same frames on film. It is significant to extract the time stamp information on the film so that the researchers can efficiently use the image data. This paper introduces an intelligent method for extracting time stamp information, namely, the Convolutional Neural Network (CNN), which is an algorithm in deep learning of multilayer neural network structures and can identify time stamp character in the scanned solar images. We carry out the time stamp decoding for the digitized data from the National Solar Observatory from 1963 to 2003. The experimental results show that the method is accurate and quick for this application. We finish the time stamp information extraction for more than 7 million images with the accuracy of 98\%.

\end{abstract}


\section{Introduction} \label{sec:intro}

The chromosphere is a layer of atmosphere between the photosphere and the corona. The chromospheric magnetic field structure is high dynamic, and the most intensive activities are solar flares. In order to study the solar flares and other solar activities, it is necessary to accumulate the flare observations in the chromosphere for many years. Therefore, a number of solar telescopes have been established around the world, for example, the Solar Magnetic Field Telescope (SMFT) \citep{AiGX1986} in Huairou, China, and McMath-Pierce Solar Telescope\footnote{\url{https://www.noao.edu/outreach/kptour/mcmath.html}} in Arizona, USA. Prior to the availability of modern digital cameras, the main medium for recording solar chromosphere data was film. In order to use rich historical data, many projects are involved to digitize historical astronomical data, and new research results are obtained from old data such as observation of a Moreton wave and wave-filament interactions associated with the renowned X9 flare on 1990 May 24 \citep{LiuR2013}, circular ribbon flares, and homologous jets \citep{WangHM2012}. Because of the huge amount of data, the time stamps of many digitized chromospheric images are still in the form that cannot be read directly by the computer, which has produced obstacles to further research. The digitization of time stamps makes the data to be more efficiently analyzed. Therefore, time stamp decoding is a significant problem that we intend to solve.

\begin{figure}[!htbp]
\centering
\includegraphics[width=0.8\textwidth]{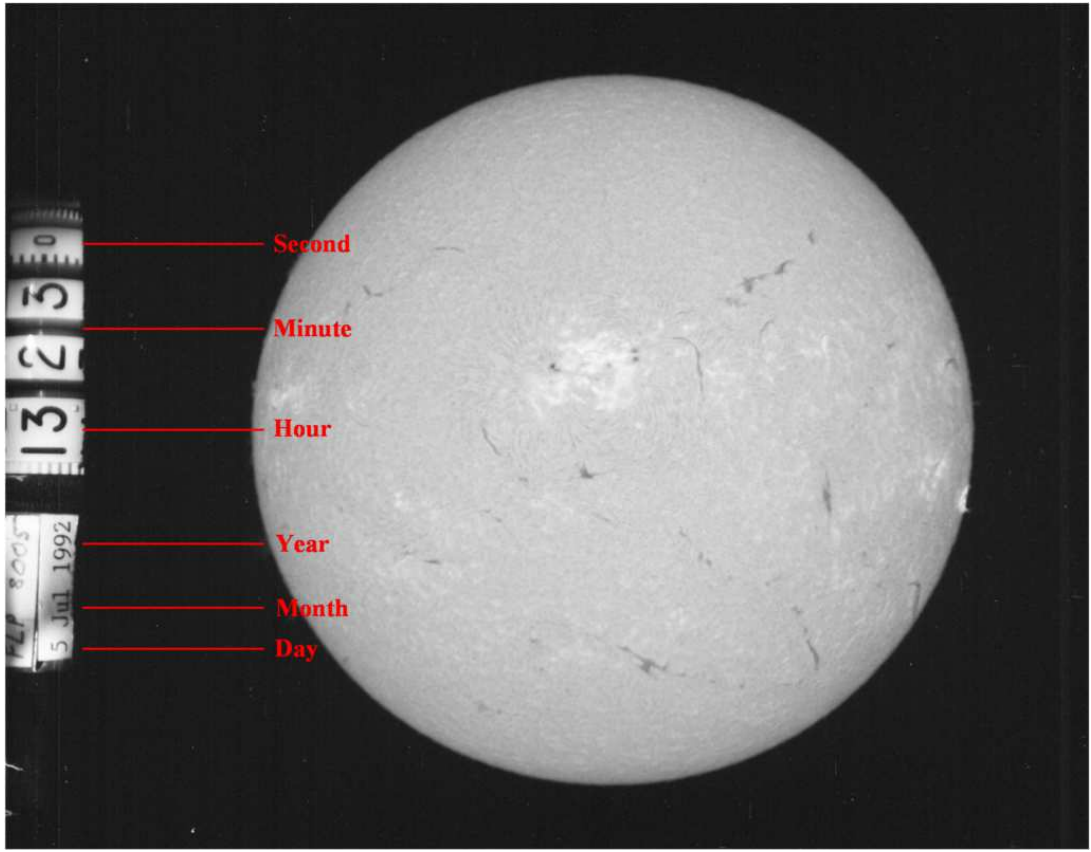}
\caption{Full-disk chromosphere image obtained in H$\alpha$ line by NSO. The time stamp is on the left side of the image, and the full-disk chromosphere image in the middle.}
\label{fig:1}
\end{figure}

From 1963 to 2003, full-disk H$\alpha$ images are recorded in 35-mm films with 1 minute or even shorter cadence at National Solar Observatory (NSO) of the US. More than 8 million pictures have been recorded and then digitized by the New Jersey Institute of Technology (NJIT), which covers hundreds of solar flares and other activities. It will create a valuable data archive of solar eruptions, which is a huge advance in solar astronomy. However, the data are useless before the decoding of time stamps. An example of a chromospheric image is shown in Figure~\ref{fig:1}. The image records some information such as the year, month, day, hour, minute, second, and film number, besides full-disk solar image. The time/date when the picture was taken is what we need to extract. As the amount of data is very large, automatically identifying the characters of the time stamp is the key to the efficient usage of the data. In order to solve the problem of character recognition, many methods have been proposed, such as support vector machine algorithm \citep{Nasien2010}, deep learning algorithm \citep{Witten2017,LeCun2015,Schmidhuber2015}, and so on.

Recently, the Convolutional Neural Network (CNN) \citep{Krizhevsky2017,Lawrence1997} is a popular deep learning algorithm with high accuracy in classification. It has been widely used in face recognition \citep{SunY2014}, image classification \citep{Krizhevsky2017}, speech recognition \citep{Hinton2012}, character recognition \citep{ZhengS2016,Goodfellow2013}, etc. \citet{ZhengS2016} applied it for character recognition in the sunspot drawings of Yunnan Observatory, with the accuracy of 98.5\%. \citet{Goodfellow2013} has applied CNN to the Street View House Numbers (SVHNs) dataset with the accuracy of 96\%. We adopt the CNN for character recognition because of the high accuracy. The selection effect of samples is the key to the recognition accuracy of the CNN. However, the characters in the time stamp are specific and not included in any digital sample database. We need to create a sample database for them as a training set. In addition, many images are ambiguous, and there is still a big hindrance to solving character segmentation and recognition.

In this paper, we present an intelligent recognition method for automatic segmentation and recognition of characters based on CNN. The paper is organized as follows. Section~\ref{sec:cnn} is an introduction to CNN algorithm. In Section~\ref{sec:stampRec}, we apply the CNN algorithm to time stamp recognition. Section~\ref{sec:resDis} demonstrates the recognition result of this method for the time stamp. Finally, we give a conclusion in Section~\ref{sec:con}.

\section{Convolutional Neural Network} \label{sec:cnn}

\begin{figure}[!htbp]
\centering
\includegraphics[width=0.8\textwidth]{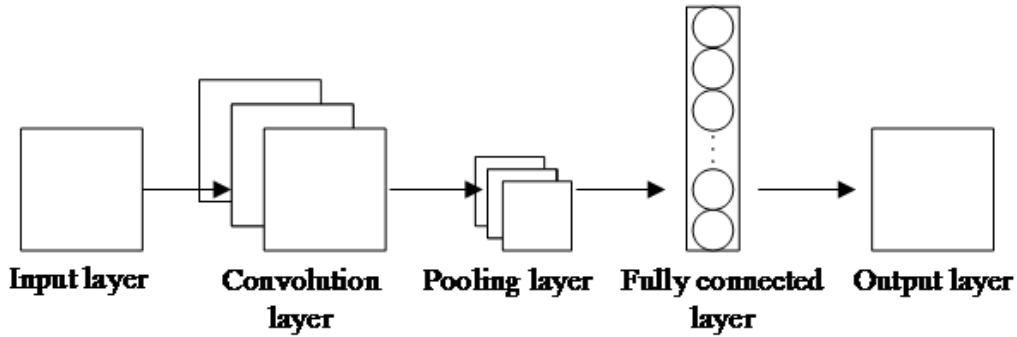}
\caption{Convolutional neural network structure, with input layer, convolution layer, pooling layer, fully connected layer, and output layer.}
\label{fig:2}
\end{figure}

The CNN \citep{Krizhevsky2017,Lawrence1997,Kim2014} includes input layer, convolutional layers, pooling layers, fully connected layers, and output layer. A typical structure is shown in Figure~\ref{fig:2}. Feature vectors in the outer layer are extracted from data in the input layer by the convolutional layer, the pooling layer, and the fully connected layer and then used in classifying the input data by logistic regression.

Multiple convolutional layers, pooling layers, and fully connected layers are possible in the CNN. The convolution layer detects the characteristics of the input layer to the maximum extent by randomly generating sufficient convolution kernels. A large number of feature maps are generated after passing through the convolutional layer. The convolution layer is usually followed by an activation function which is used to convert features from a linear space into a nonlinear space to achieve the nonlinear classification \citep{GuJX2018}. ReLu, sigmod, and tanh are commonly applied as the activation functions. In this paper, ReLu is adopted, which can effectively prevent overfitting problems. The pooling layer is a feature filter for the convolutional layer to preserve the main features and to reduce the amount of computation. It is often placed in the middle of two convolutional layers.

The data, which processed by multiple convolution layers and pooling layers, are connected to one or more fully connected layers. In a fully connected layer, each neuron is connected to all neurons in the upper layer to combine the features extracted previously, so the extracted features can be completely preserved and unaffected by the position in the original image. The output value of the output layers is classified by logistic regression. Softmax regression is usually used when dealing with multiclassification problems. The Softmax regression outputs the probability value of the sample for each class and selects the class corresponding to the maximum probability as the recognition result of the sample. In addition, the recognition accuracy of CNN is closely related to the quality and quantity of samples. The richer the training samples, the higher the recognition accuracy.

\section{Time Stamp Character Recognition Based on CNN} \label{sec:stampRec}

\begin{figure}[!htbp]
\centering
\includegraphics[width=0.6\textwidth]{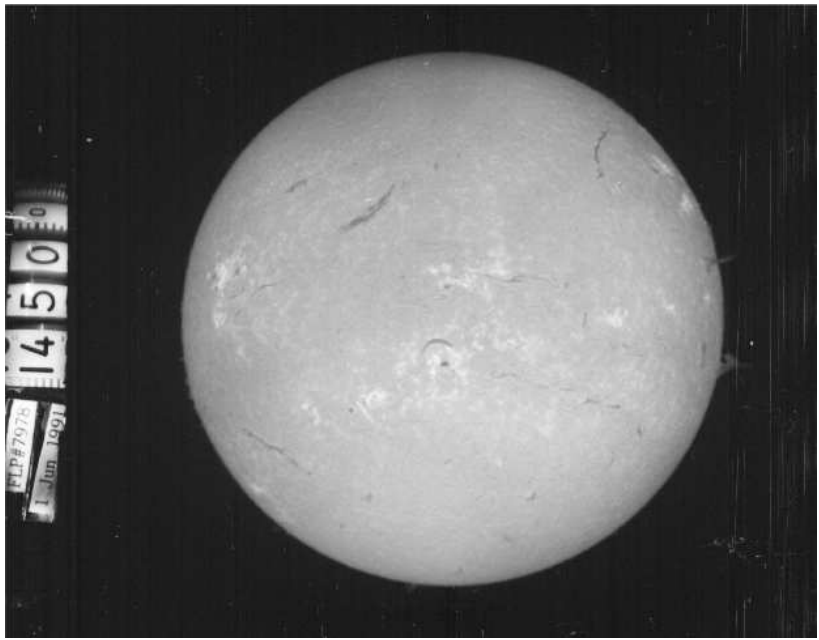}
\caption{Black character image.}
\label{fig:3}
\end{figure}

\begin{figure}[!htbp]
\centering
\includegraphics[width=0.6\textwidth]{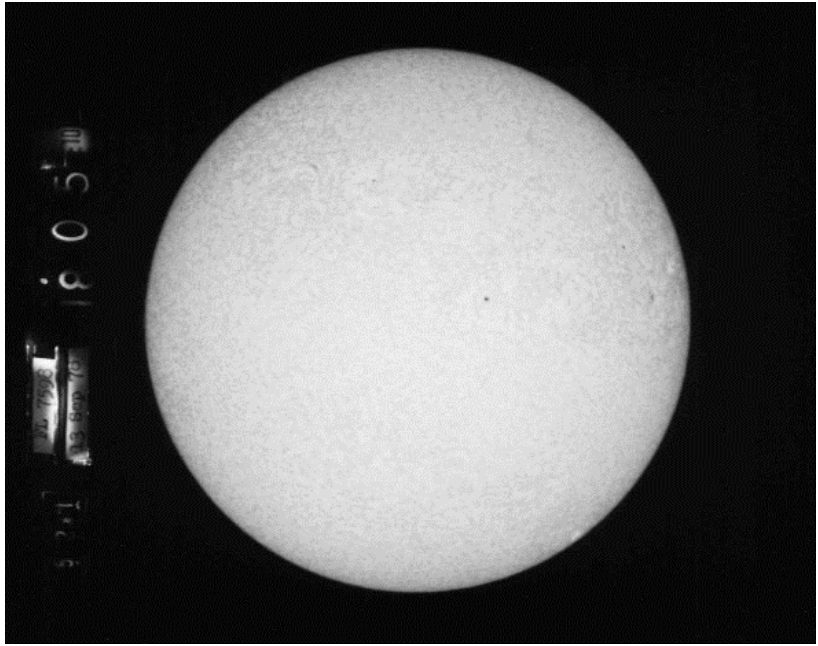}
\caption{White character image.}
\label{fig:4}
\end{figure}

\begin{figure}[!htbp]
\centering
\includegraphics[width=0.9\textwidth]{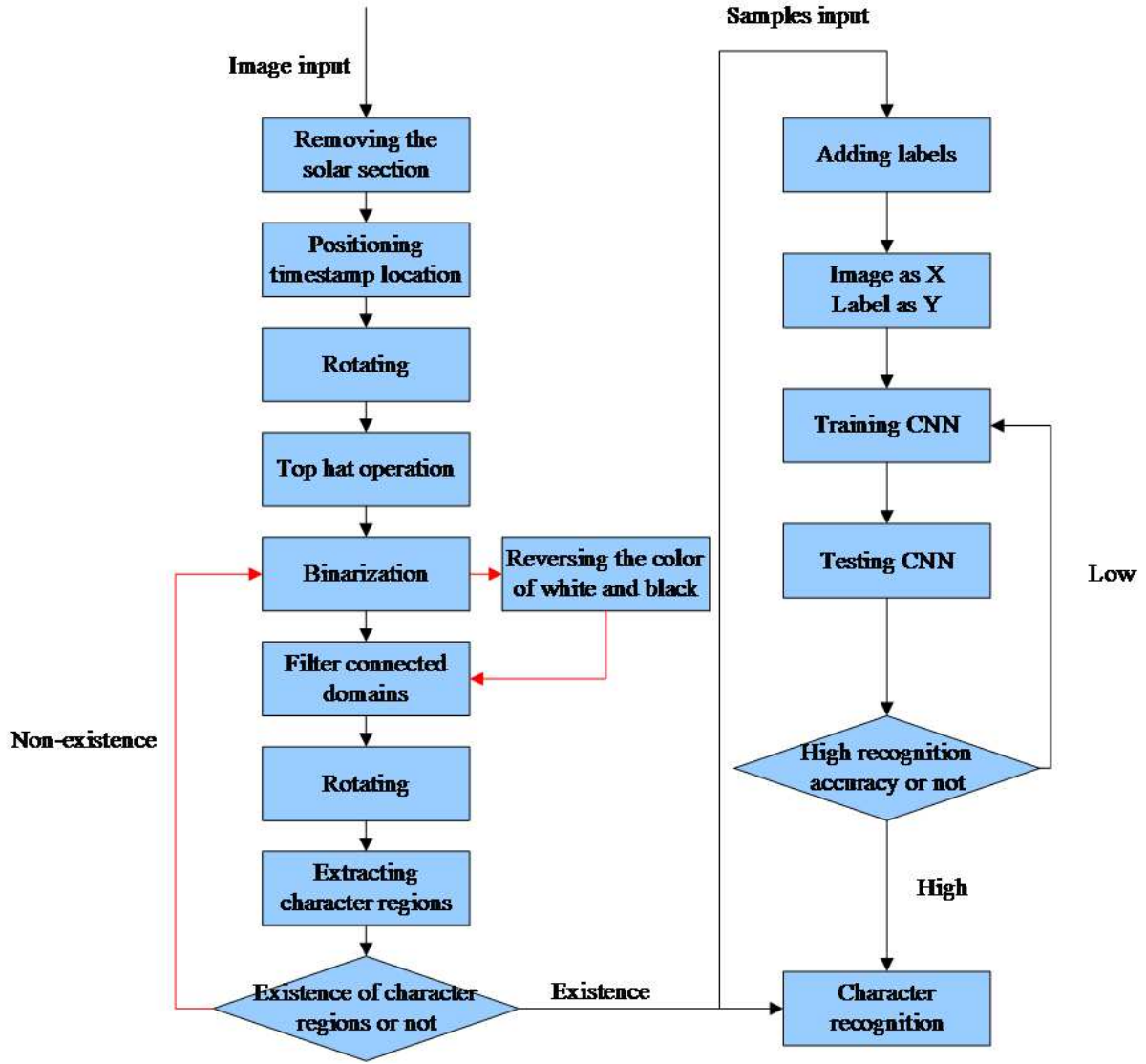}
\caption{Algorithm flow chart.}
\label{fig:5}
\end{figure}

The information of year, month, day, hour, and minute is what we need to extract in the image. Figures \ref{fig:3} and \ref{fig:4} show chromospheric pictures with two types of the time stamp. The time stamp in Figure~\ref{fig:3} is black on white, while Figure~\ref{fig:4} is white on black. The time stamps are uneven, the format and color of the characters are inconsistent, the YMD (year, month, and day) characters are small, and the characters in many pictures are illegible and difficult to recognize. However, the date information is continuous, and there are many images on the same date. So, we only need to get the date of the first picture every day without intelligent recognition. That part was achieved manually. The CNN is used in identifying the HM (hour and minute) characters. The flow chart of the CNN algorithm for recognizing time stamp characters is shown in Figure~\ref{fig:5}. It consists of two independent parts: one for character segmentation (Section~\ref{subsec:charSeg}) and the other for character recognition by CNN (Section~\ref{subsec:charRec}).

The left part of the flow chart introduces image segmentation, and the right part introduces character recognition. The input image is processed by white characters by default. If no character areas can be extracted, the process returns to the binarization. Reverse the color of white and black in the binary image; The CNN will be retrained when it has a low recognition rate for the test samples.

\subsection{Characters Segmentation} \label{subsec:charSeg}

The size of the original image is $1600\times 2048$, as shown in Figure~\ref{fig:3}. The time stamp is on the left or right side of the picture and the character format is different. Characters are divided into two categories, one is black and the other is white, which need to be dealt with separately. The character segmentation steps are as follows.

\begin{figure}[!htbp]
\gridline{\fig{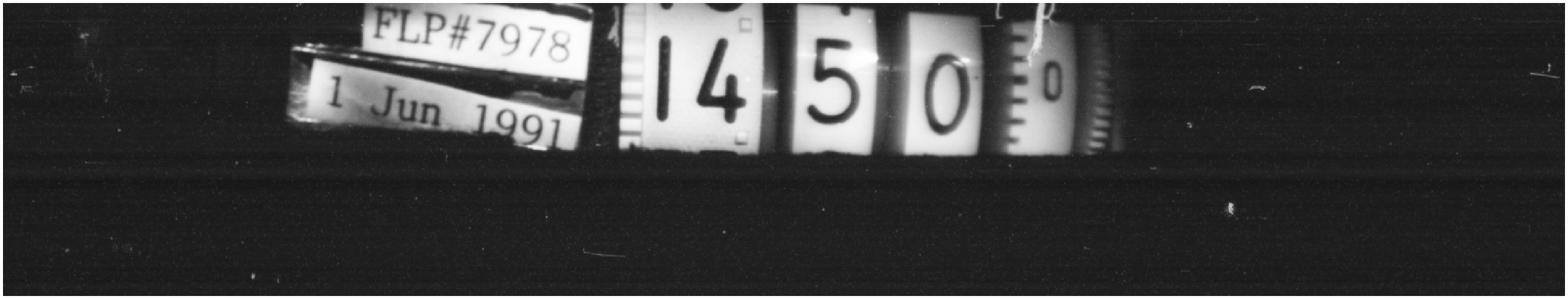}{0.35\textwidth}{(a)}
          \fig{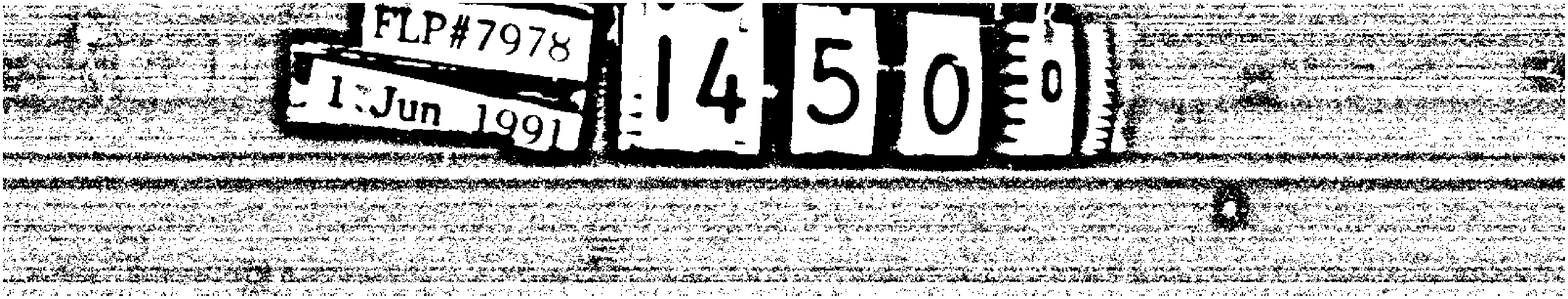}{0.35\textwidth}{(b)}
          }
\gridline{\fig{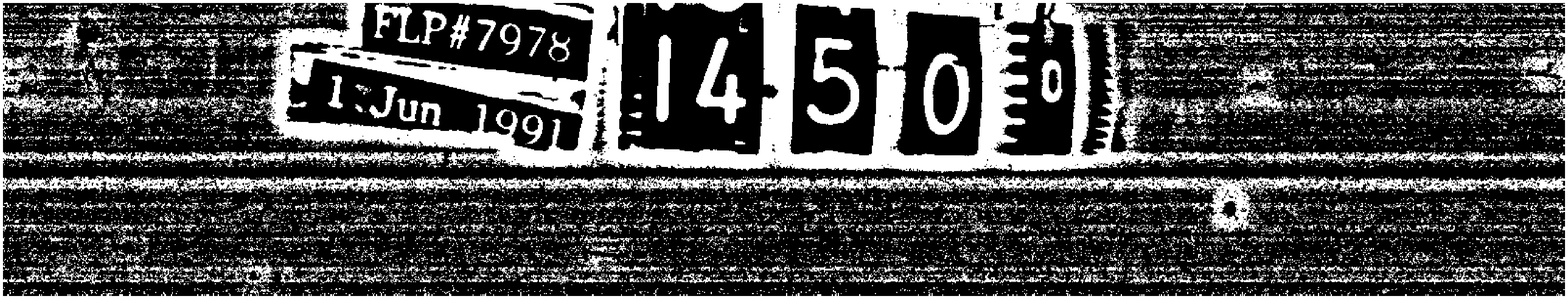}{0.35\textwidth}{(c)}
          \fig{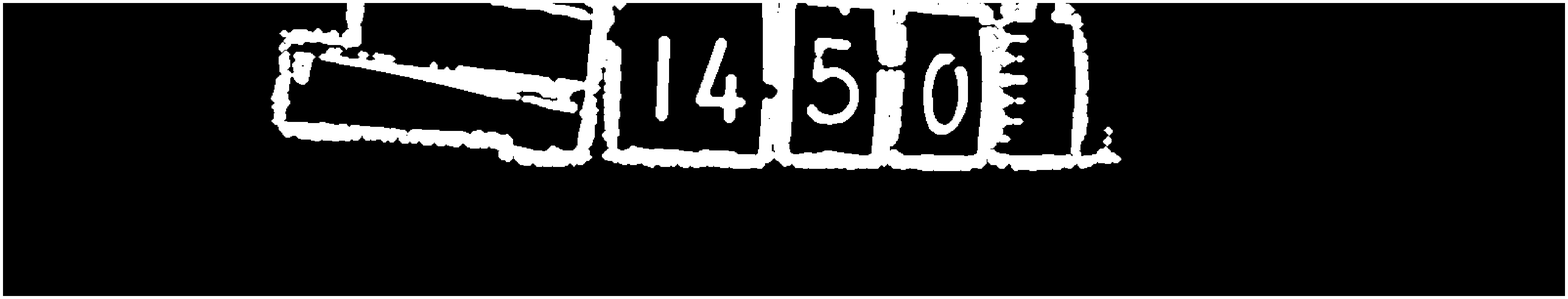}{0.35\textwidth}{(d)}
          }
\gridline{\fig{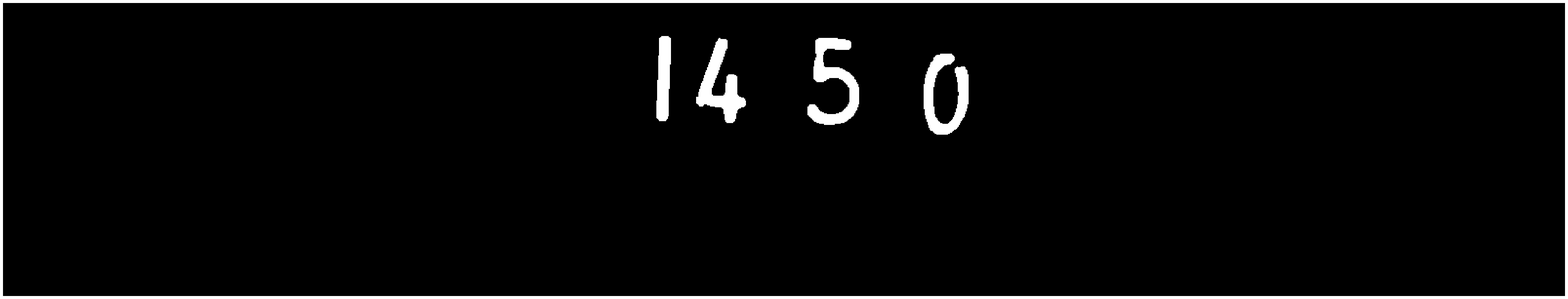}{0.35\textwidth}{(e)}
          \fig{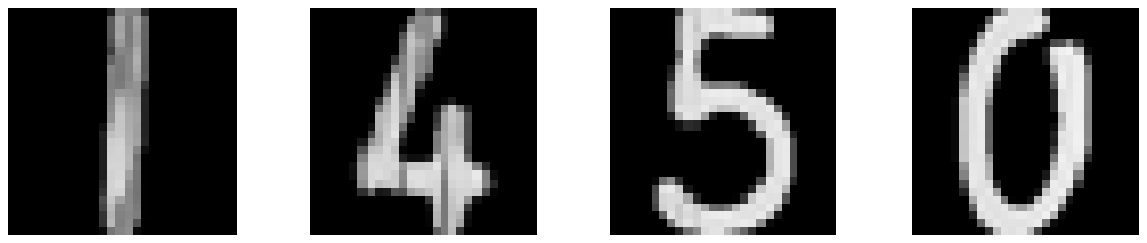}{0.35\textwidth}{(f)}
          }
\caption{Extract characters from Figure~\ref{fig:3}: (a) position the time stamp and rotate, (b) after top hat operation and binarization, (c) black and white inversion of binary image, (d) extract related connectivity domain, (e) extract character regions, and (f) characters extraction result.}
\label{fig:6}
\end{figure}

\begin{quotation}
\textit{Step 1}. Remove the part of the solar disk from the picture and obtain the left and right sides of the picture. 

\textit{Step 2}. Get the picture with a time stamp based on the intensity variance across the picture, and rotate the picture to adjust the direction of the characters (Figure~\ref{fig:6}(a)). 

\textit{Step 3}. Eliminate noise in pictures with top hat operation (Figure~\ref{fig:6}(b)). 

\textit{Step 4}. Binarize the picture by the Sauvola algorithm \citep{Sauvola2000}. 

\textit{Step 5}. Reserve connectivity domain of which area is in $(500,1000)$. 

\textit{Step 6}. Extract character regions using stroke width transform algorithm \citep{Epshtein2010,LiY2012} (Figure~\ref{fig:6}(e)). 

\textit{Step 7}. If there are no character regions, return to Step 4. Reverse the color of white and black in the binary image which is obtained in Step 4 (Figure~\ref{fig:6}(c)) to get white characters, to allow black characters in the original image to be extracted. This ensures data consistency so that the next steps are as identical as possible. After Step 5 as shown in Figure~\ref{fig:6}(d). If there are still no character regions after the image is reversed, it means that there are no characters in the current picture. Because there are only two forms of time stamp and few a part of the images that do not contain time stamps, the time stamp characters cannot be extracted from these images during the above process. 

\textit{Step 8}. Extract the corresponding region from the original image according to the binary image, and resize each of the characters to $28 \times 28$ (Figure~\ref{fig:6}(f)).
\end{quotation}

\subsection{Characters Recognition} \label{subsec:charRec}

\begin{figure}[!htbp]
\centering
\includegraphics[width=0.96\textwidth]{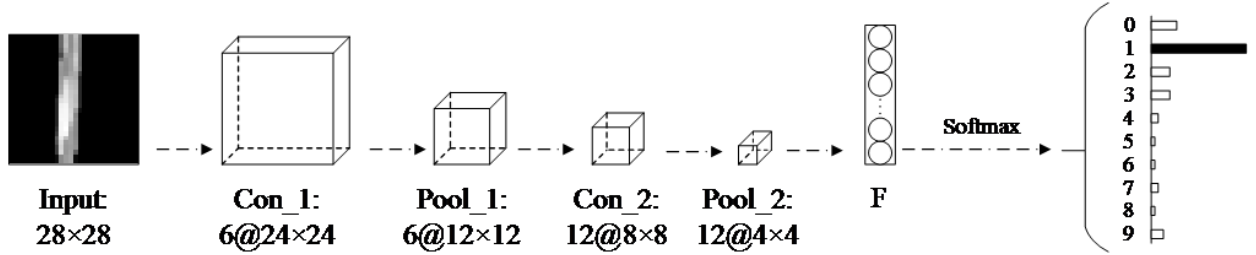}
\caption{Convolution neural network structure of character recognition, included a input layer, two convolution layers, two pooling layers, and a fully connected layer.}
\label{fig:7}
\end{figure}

The CNN model for time stamp character recognition consists of two convolutional layers, two pooling layers, and a fully connected layer (Figure~\ref{fig:7}). In the first convolutional layer Con\underline{~}1, 6 different convolutional kernels of size $5\times 5$ are used to take convolution operation on character pictures with the size of $28\times 28$. After Con\underline{~}1, the original character picture becomes a $24\times 24\times 6$ feature map. The first pooling layer Pool\underline{~}1 filters the feature map using maximum pooling function with the sliding window of $2\times 2$. Then, it becomes a feature map of size $12\times 12\times 6$. The convolutional layer Con\underline{~}2 contains 10 kernels of $5\times 5$. The pooling layer Pool\underline{~}2 does the same as Pool\underline{~}1. These feature maps are taken as the inputs into the fully connected layer to obtain the feature vector. Finally, the vector is classified by the Softmax function to obtain the recognition result.

The training steps of the CNN in this paper are divided into the following three steps.

\begin{quotation}
\textit{Step 1}. Add labels to the single-character images as samples for training the network. 

\textit{Step 2}. The character image is used as the $X$ vector for the input layer, and the label of the image is used as the $Y$ vector. 

\textit{Step 3}. The network is trained by forwarding propagation and back propagation algorithms, and its coefficients are updated by loop iteration. A network structure with higher recognition accuracy is obtained in the end.
\end{quotation}

To train the CNN, we select 100,000 single-character images of size $28 \times 28$, which were cut from the original images with white characters, as training samples, 10,000 samples per character. These characters are recognizable by humans and labeled manually. There is no need to deal with time stamps unrecognizable by humans, because it is impossible to verify the recognition correctness. 9000 images are randomly selected as samples to train the network. The remaining samples are used as testing samples to test the recognition accuracy of the network. The test results are shown in Table~\ref{tab:1}. From the table, the recognition accuracy of each character is over 98\%, and it takes only about 6 seconds to recognize 1000 pictures.

\begin{table}[!htbp]
\centering
\caption{Test results of CNN identification.}
\label{tab:1}
\begin{tabular*}{0.86\textwidth}{lcccc}
\hline
\hline
Character & Total numbers & Recognition errors & Recognition accuracy rate & Time cost (s)  \\
\hline
0 & 1000 & 20 & 0.980 & 6.01  \\
1 & 1000 & 4 & 0.996 & 6.11  \\
2 & 1000 & 8 & 0.992 & 5.96  \\
3 & 1000 & 3 & 0.997 & 6.28  \\
4 & 1000 & 1 & 0.999 & 6.24  \\
5 & 1000 & 19 & 0.981 & 6.52  \\
6 & 1000 & 12 & 0.988 & 6.05  \\
7 & 1000 & 3 & 0.997 & 6.17  \\
8 & 1000 & 12 & 0.988 & 5.95  \\
9 & 1000 & 13 & 0.987 & 6.11  \\
\hline
\end{tabular*}
\end{table}

\begin{figure}[!htbp]
\centering
\includegraphics[width=0.8\textwidth]{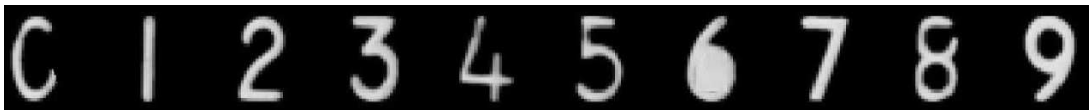}
\caption{Character samples affected by illumination and background noise.}
\label{fig:8}
\end{figure}

At present, the commonly used methods for character recognition are the Optical Character Recognition (OCR) \citep{Mohiuddin1999} and character recognition based on deep neural network. It is well known that OCR recognizes standard characters effectively. So we did an experiment based on open recognition engine TESSERACT \citep{Smith2007}. We train it in the same way as CNN, and the same way to test it. The test results are shown in Table~\ref{tab:2} that the highest recognition accuracy is 96.8\% and the lowest is 93.2\% and the lowest time cost of testing 1000 samples is 8.23 seconds. The CNNs, on the contrary, have higher recognition accuracy and lower time cost. The reason for the relatively low recognition accuracy of OCR is that characters extracted from time stamps are affected by some interference, such as illumination interference, background noise interference, as shown in Figure~\ref{fig:8}. It is hard for OCR to handle these situations. So it can be concluded from the comparative experiments that CNN has better robustness, stronger antijamming, and lower time consumption than OCR.

\begin{table}[!htbp]
\centering
\caption{Test results of TESSERACT.}
\label{tab:2}
\begin{tabular*}{0.86\textwidth}{lcccc}
\hline
\hline
Character & Total numbers & Recognition errors & Recognition accuracy rate & Time cost (s)  \\
\hline
0 & 1000 & 57 & 0.943 & 8.95  \\
1 & 1000 & 45 & 0.955 & 8.32  \\
2 & 1000 & 53 & 0.947 & 8.28  \\
3 & 1000 & 50 & 0.950 & 8.84  \\
4 & 1000 & 59 & 0.941 & 8.23  \\
5 & 1000 & 40 & 0.960 & 8.44  \\
6 & 1000 & 49 & 0.951 & 8.52  \\
7 & 1000 & 32 & 0.968 & 8.39  \\
8 & 1000 & 68 & 0.932 & 8.45  \\
9 & 1000 & 60 & 0.940 & 8.69  \\
\hline
\end{tabular*}
\end{table}

\subsection{Date Check} \label{subsec:dataChk}

\begin{figure}[!htbp]
\centering
\includegraphics[width=0.7\textwidth]{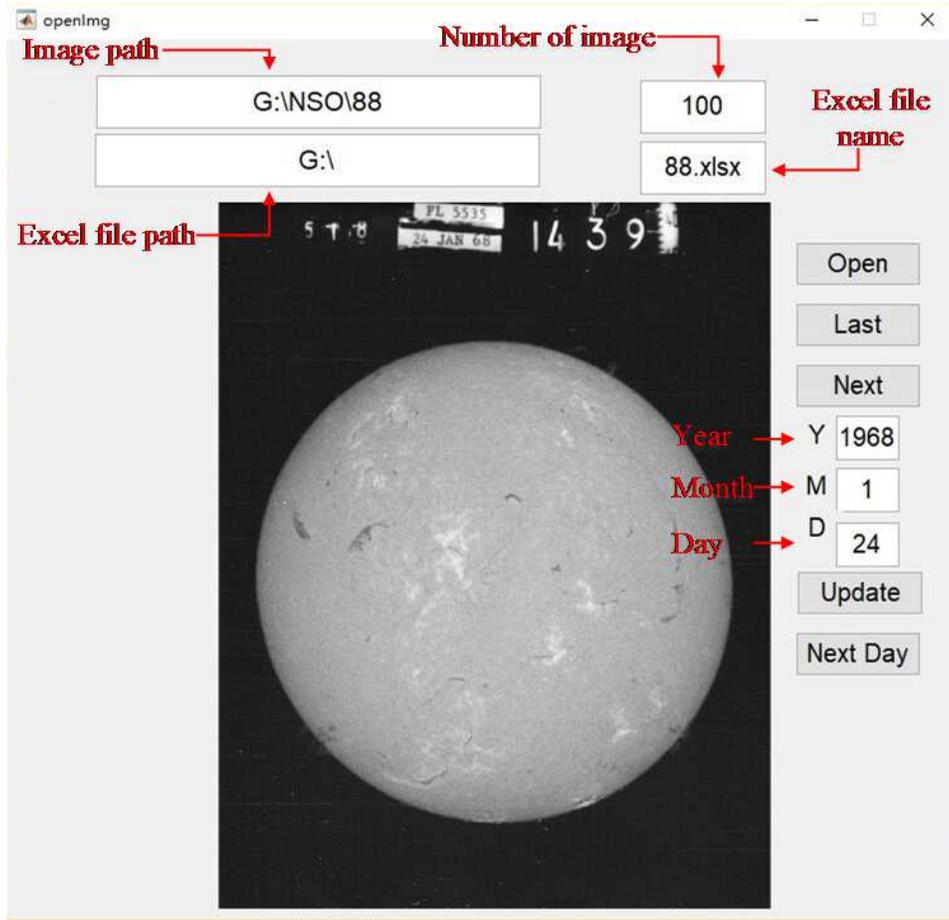}
\caption{Graphic interface of date check.}
\label{fig:9}
\end{figure}

After the hour and minute in the time stamp are identified, another important step is to complete the information of the date (year, month, and day). Since the date of the photo may not be continuous and cannot be filled in automatically by the program, it is necessary to confirm the date manually. Although the dates are not continuous, they are all in order, and the volume number, which is recorded in the folder name, of the film helps to determine the range of date. In addition, the photographing time is mostly continuous and the 24-hour timekeeping method is used; it is easy to judge whether the date has changed. For example, if time information of the first picture is ``2359'' and that of the second picture ``000'', the date information of the second picture can be added one day based on the first picture. So for images over a period of time, it only needs to know the observation date of the first picture. However, some dates are not continuous, so a manual check is required. So we adopt a user graphical interface (Figure~\ref{fig:9}) to assist in the date confirmation. Only the first few pictures of a day need to be verified. If the date is incorrect, modify it manually, and the program will automatically update all dates in the subsequent pictures.

Fill in the paths of the original image and the record table in the corresponding text box of the program. Click on the ``Open'' button to open the first image in the folder and its date information is displayed in the corresponding text box. Click on the ``Next'' or ``Last'' button to open the next or previous image, respectively. Click on the ``Update'' button to update the date. The ``Next day'' button is used to jump directly to the next day. Finally, the updated contents are saved in the corresponding files.

\section{Result and Discussion} \label{sec:resDis}

To further test the recognition accuracy of the network under actual conditions, we randomly selected 10,000 original images for testing. Table~\ref{tab:3} shows the accuracy of the testing results recognized by CNN, which is confirmed manually. Misrecognizing 1 character occurs 202 times, misrecognizing 2 characters occurs 10 times, and no situation occurs for misrecognizing more than 3 characters simultaneously. The recognition accuracy rate is 97.9\%, and the average time taken for each picture is 0.09 seconds. The statistics of recognition results for each character are shown in Table~\ref{tab:4}.

\begin{table}[!htbp] \addtolength{\tabcolsep}{-1pt}
\centering
\caption{Statistical table of recognition results.}
\label{tab:3}
\begin{tabular*}{\textwidth}{lccccccc}
\hline
\hline
 & Correct & 1 error & 2 errors & 3 errors & 4 errors & Average time cost (s) & Recognition accuracy rate  \\
\hline
Numbers & 9788 & 202 & 10 & 0 & 0 & 0.09 & 97.9\%  \\
\hline
\end{tabular*}
\end{table}

\begin{table}[!htbp]
\centering
\caption{Confusion matrix of character recognition results.}
\label{tab:4}
\begin{tabular*}{0.805\textwidth}{lcccccccccc}
\hline
\hline
Character & 0 & 1 & 2 & 3 & 4 & 5 & 6 & 7 & 8 & 9  \\
\hline
0 & 4558 & 0 & 2 & 108 & 0 & 0 & 16 & 0 & 0 & 1  \\
1 & 0 & 11147 & 0 & 0 & 0 & 0 & 0 & 0 & 0 & 0  \\
2 & 0 & 0 & 4826 & 0 & 0 & 0 & 0 & 0 & 0 & 0  \\
3 & 0 & 1 & 85 & 3890 & 0 & 1 & 0 & 0 & 0 & 0  \\
4 & 0 & 0 & 0 & 0 & 4176 & 2 & 0 & 0 & 0 & 0  \\
5 & 0 & 0 & 0 & 0 & 0 & 4362 & 0 & 0 & 0 & 0  \\
6 & 0 & 0 & 0 & 0 & 0 & 0 & 1299 & 0 & 0 & 0  \\
7 & 0 & 0 & 0 & 0 & 0 & 0 & 0 & 1918 & 0 & 0  \\
8 & 0 & 0 & 0 & 0 & 0 & 0 & 0 & 1 & 1846 & 0  \\
9 & 0 & 0 & 0 & 0 & 0 & 0 & 0 & 0 & 1 & 1760  \\
\hline
Total & 4558 & 11148 & 4913 & 3998 & 4176 & 4365 & 1315 & 1919 & 1847 & 1761  \\
Number of erros & 0 & 1 & 87 & 108 & 0 & 3 & 16 & 1 & 1 & 1  \\
Recognition rate & 1 & 0.99 & 0.98 & 0.97 & 1 & 0.99 & 0.99 & 0.99 & 0.99 & 0.99  \\
\hline
\end{tabular*}
\end{table}

\begin{figure}[!htbp]
\gridline{\fig{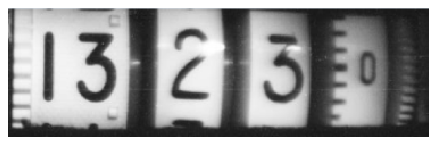}{0.25\textwidth}{(a)}
          \fig{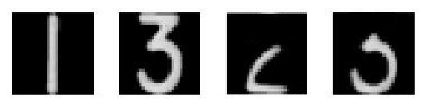}{0.3\textwidth}{(b)}
          }
\gridline{\fig{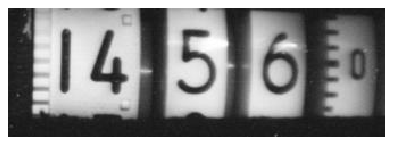}{0.25\textwidth}{(c)}
          \fig{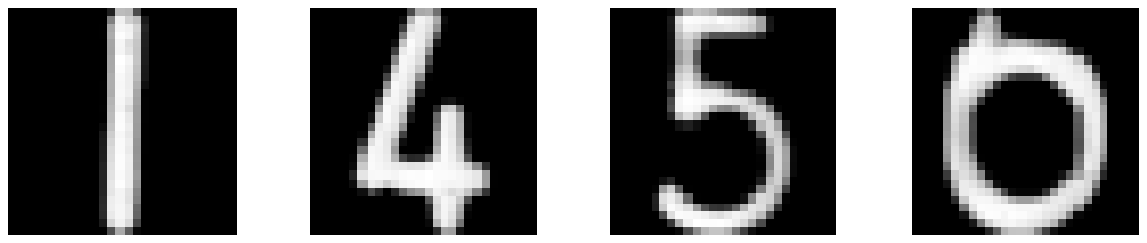}{0.3\textwidth}{(d)}
          }
\caption{Original image (a, c) and segmentation result (b, d). As shown in figure (b, d), the third character and the fourth character are not completely split in (b), the fourth character is not completely split in (d), resulting in incorrect recognition results.}
\label{fig:10}
\end{figure}

\begin{figure}[!htbp]
\centering
\gridline{\fig{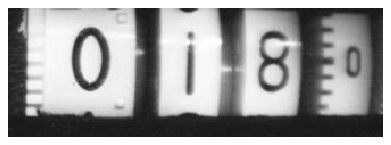}{0.25\textwidth}{(a)}
          \fig{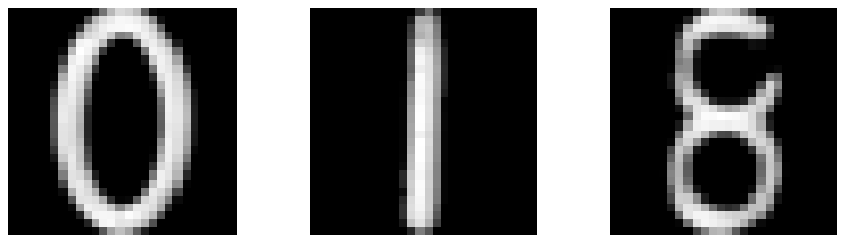}{0.35\textwidth}{(b)}
          }
\gridline{\fig{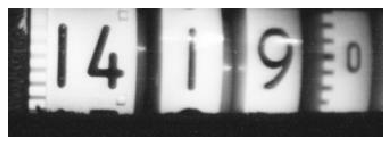}{0.25\textwidth}{(c)}
          \fig{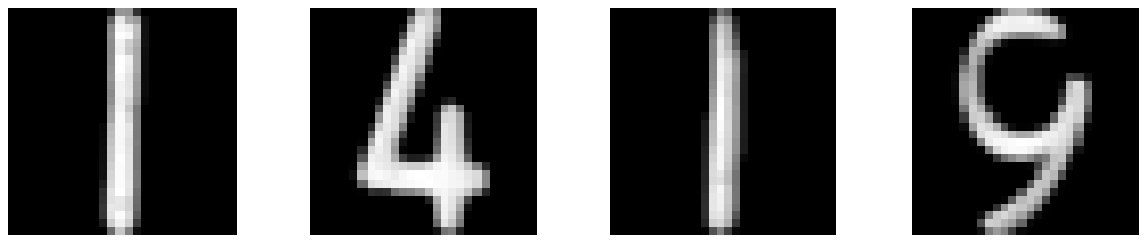}{0.35\textwidth}{(d)}
          }
\caption{Original image (a, c) and segmentation result (b, d), respectively. As shown in figure (b, d), the recognition result is correct, even though the fourth character is partially split.}
\label{fig:11}
\end{figure}

Table~\ref{tab:4} shows that the recognition accuracy of the character ``0'' is 100\%, that of ``1'', ``5'', and ``7'' is greater than 99.9\%, and that of other characters is above 97.3\%. The average recognition accuracy of all the characters is 99.5\%. However, the recognition error rates of the characters ``2'', ``3'', and ``6'' are higher, mainly due to these characters being affected by the light, as shown in Figure~\ref{fig:10}. When they are affected by illumination, they are easily destroyed by the local binary algorithm leading to structural breaks. The character fragments are considered to be noise in the next step of the algorithm because of their small area, which will affect the recognition results (e.g., Figures \ref{fig:10}(b) and \ref{fig:10}(d). However, these images affected by lighting only account for a small part of the whole samples, as shown in Table~\ref{tab:4}, so they contribute a little to the average recognition accuracy. Besides, the recognition results of some characters are not affected by illumination, such as ``8'' and ``9'', as shown in Figure~\ref{fig:11}. When there is lighting interference on the images, their main structures are preserved so that their recognition results are not affected. The defective structures can be identified, which is one of the advantages of CNN.

Although these images affected by lighting only account for a small part, to solve this problem, our further plan is adding some samples affected by lighting to the training set and improving the algorithm of character segmentation.

\begin{figure}[!htbp]
\centering
\includegraphics[width=0.9\textwidth]{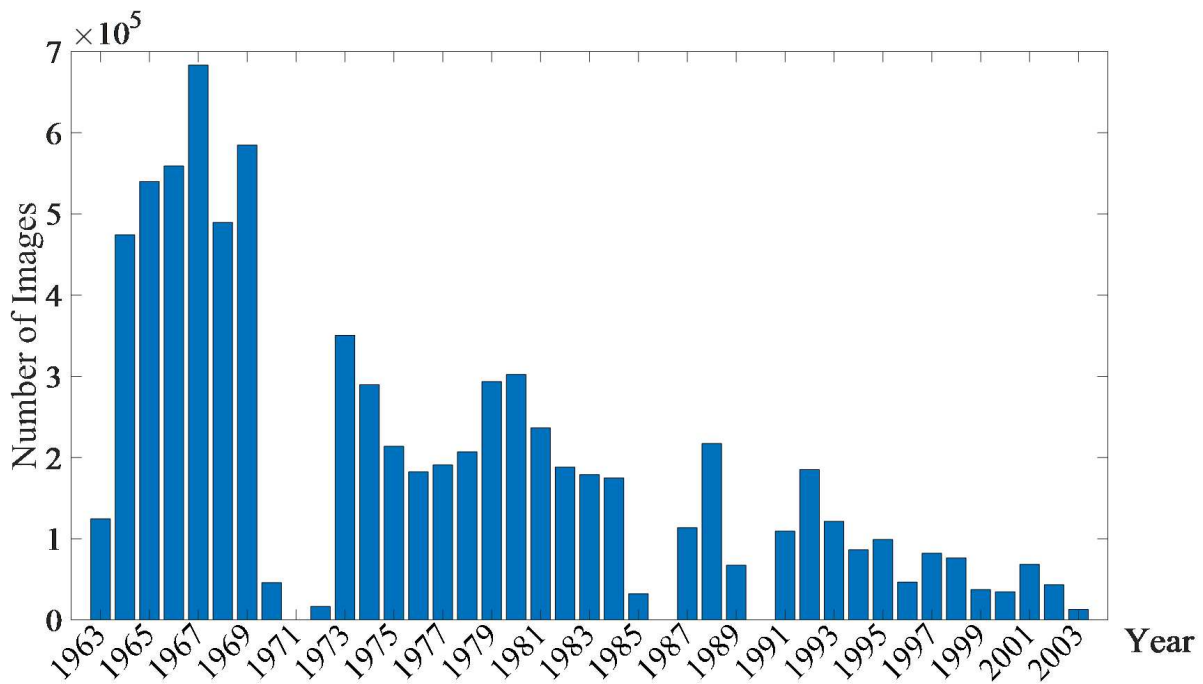}
\caption{Number of pictures per year.}
\label{fig:12}
\end{figure}

In total, we get date/time information for more than 7 million pictures of 38 years, as shown in Table~\ref{tab:5}. The remaining unprocessed images such as those in 1971, 1986, and 1990 are due to time stamps that are beyond human recognition or do not have time stamps, about 10\% of the total. It is not necessary to deal with these pictures because it is impossible to verify whether they are recognized correctly or not. The number of pictures per year is also shown in a bar chart as shown in Figure~\ref{fig:12}. The number of pictures rose slowly from 1963 to 1967, peaking in 1967 with about 700 thousand pictures. After 1967, the number of pictures declined dramatically. In 2003, there were about 13,000 pictures.

\begin{table}[!htbp] \addtolength{\tabcolsep}{+30pt}
\centering
\caption{Annual total amount of images.}
\label{tab:5}
\begin{tabular}{lr}
\hline
\hline
Year & Number  \\
\hline
1963 & 124469  \\
1964 & 474370  \\
1965 & 539969  \\
1966 & 559126  \\
1967 & 683452  \\
1968 & 489625  \\
1969 & 584851  \\
1970 & 46047  \\
1971 & 0  \\
1972 & 16624  \\
1973 & 350606  \\
1974 & 289724  \\
1975 & 213868  \\
1976 & 182321  \\
1977 & 190870  \\
1978 & 207012  \\
1979 & 293451  \\
1980 & 302258  \\
1981 & 236533  \\
1982 & 188228  \\
1983 & 178873  \\
1984 & 174902  \\
1985 & 32324  \\
1986 & 0  \\
1987 & 113652  \\
1988 & 217226  \\
1989 & 67403  \\
1990 & 0  \\
1991 & 109306  \\
1992 & 185144  \\
1993 & 121401  \\
1994 & 86425  \\
1995 & 98994  \\
1996 & 46638  \\
1997 & 82101  \\
1998 & 76434  \\
1999 & 37270  \\
2000 & 34679  \\
2001 & 68460  \\
2002 & 43292  \\
2003 & 12983  \\
\hline
Total & 7760911  \\
\hline
\end{tabular}
\end{table}

\section{Conclusion} \label{sec:con}

In this paper, we describe an intelligent algorithm to extract the time stamp from traditional films based on CNN. The experimental results show that the method has a good result and meets the speed and quality requirements for identification. It also has strong portability in solving the same type of problems in similar applications.

Finally, we get date/time information for more than 7 million pictures which are recorded by NSO of the US. This greatly reduces the amount of manual work, so that this batch of data can be effectively utilized by researchers as soon as possible. The method proposed in this paper can be applied to character recognition in other historical image, such as handwritten character recognition in sunspot drawing.

\acknowledgements

This work is supported in part by the National Natural Science Foundation of China under Grants U1731124, U1531247, 11427803, 11427901, and 11873062, the 13th Five-year Informatization Plan of Chinese Academy of Sciences under Grant XXH13505-04, and the Beijing Municipal Science and Technology Project under Grant Z181100002918004. Haimin Wang acknowledges the support of US NSF under grant AGS-1620875. The authors are grateful to the National Solar Observatory for providing the original film data.

%\bibliographystyle{aasjournal}
%\bibliography{zjf18}{}

\end{document}